\documentclass[11pt,a4paper]{article}
\usepackage{graphicx}
\usepackage{jcappub}

\def\sun{\hbox{$\odot$}}

\def\la{\mathrel{\mathchoice {\vcenter{\offinterlineskip\halign{\hfil  
$\displaystyle##$\hfil\cr<\cr\sim\cr}}}  
{\vcenter{\offinterlineskip\halign{\hfil$\textstyle##$\hfil\cr  
<\cr\sim\cr}}}  
{\vcenter{\offinterlineskip\halign{\hfil$\scriptstyle##$\hfil\cr  
<\cr\sim\cr}}}  
{\vcenter{\offinterlineskip\halign{\hfil$\scriptscriptstyle##$\hfil\cr  
<\cr\sim\cr}}}}}  
\def\ga{\mathrel{\mathchoice {\vcenter{\offinterlineskip\halign{\hfil  
$\displaystyle##$\hfil\cr>\cr\sim\cr}}}  
{\vcenter{\offinterlineskip\halign{\hfil$\textstyle##$\hfil\cr  
>\cr\sim\cr}}}  
{\vcenter{\offinterlineskip\halign{\hfil$\scriptstyle##$\hfil\cr  
>\cr\sim\cr}}}  
{\vcenter{\offinterlineskip\halign{\hfil$\scriptscriptstyle##$\hfil\cr  
>\cr\sim\cr}}}}}

\title{An independent limit on the axion mass from the variable white dwarf  
star R548}  
  
\author[a,b]{A. H. C\'orsico,}  
\author[a,b]{L. G. Althaus,}  
\author[c]{A. D. Romero,}  
\author[d]{A. S. Mukadam,}  
\author[e,f]{E. Garc\'{\i}a--Berro,}  
\author[g,f]{J. Isern,}  
\author[c]{S. O. Kepler,}  
\author[h,a]{M. A. Corti}  
  
\affiliation[a]{Facultad de Ciencias Astron\'omicas y Geof\'{\i}sicas,  
                Universidad  Nacional de La Plata,  
                Paseo del  Bosque s/n,  
               (1900) La Plata,  
                Argentina}  
\affiliation[b]{Instituto de Astrof\'{\i}sica La Plata,  
                CONICET-UNLP,  
                Argentina}  
\affiliation[c]{Departamento de Astronomia,  
                Universidade Federal do Rio Grande do Sul,  
                Av. Bento Goncalves 9500,  
                Porto Alegre 91501-970, RS,  
                Brazil}  
\affiliation[d]{Department of Astronomy,  
                University of Washington,  
                Seattle, WA 98195-1580,  
                USA}  
\affiliation[e]{Departament de F\'\i sica Aplicada,  
                Universitat Polit\`ecnica de Catalunya,  
                c/Esteve Terrades~5,  
                08860 Castelldefels,  
                Spain}  
\affiliation[f]{Institute for Space Studies of Catalonia,  
                c/Gran Capit\`a 2--4, Edif. Nexus 104,  
                08034  Barcelona,  
	        Spain}  
\affiliation[g]{Institut de Ci\`encies de l'Espai (CSIC),  
                Campus UAB, 08193 Bellaterra,  
   	        Spain}  
\affiliation[h]{Instituto Argentino de Radioastronom\'{\i}a  
               (CCT-La Plata, CONICET), C.C. No. 5, 1894  
                Villa Elisa, Argentina}  
  
\emailAdd{acorsico@fcaglp.unlp.edu.ar}  
  
\abstract{Pulsating white dwarfs  with hydrogen-rich atmospheres, also
  known as  DAV stars,  can be used  as astrophysical  laboratories to
  constrain  the  properties  of  fundamental particles  like  axions.
  Comparing  the  measured  cooling  rates  of these  stars  with  the
  expected  values from  theoretical models  allows us  to  search for
  sources  of additional  cooling due  to the  emission of
  weakly  interacting  particles.   In   this  paper,  we  present  an
  independent  inference of  the mass  of the  axion using  the recent
  determination  of the evolutionary  cooling rate of  R548, the  DAV class
  prototype.   We employ a  state-of-the-art code  which allows  us to
  perform   a  detailed   asteroseismological  fit   based   on  fully
  evolutionary sequences.  Stellar cooling  is the solely responsible of
  the  rates of change  of period  with time  ($\dot\Pi$) for  the DAV
  class.   Thus, the inclusion  of axion  emission in  these sequences
  notably  influences  the   evolutionary  timescales,  and  also  the
  expected pulsational properties of the DAV stars.  This allows us to
  compare  the  theoretical  $\dot\Pi$  values  to  the  corresponding
  empirical rate of change of period  with time of R548 to discern the
  presence of axion  cooling. We found that if  the dominant period at
  213.13~s in R548 is associated  with a pulsation mode trapped in the
  hydrogen envelope,  our models indicate the  existence of additional
  cooling  in this pulsating  white dwarf,  consistent with  axions of
  mass  $m_{\rm  a}  \cos^2   \beta  \sim  17.1$~meV  at  a  2$\sigma$
  confidence level.  This determination is in agreement with the value
  inferred from  another well-studied  DAV, G117$-$B15A.  We  now have
  two independent  and consistent estimates  of the mass of  the axion
  obtained from  DAVs, although additional studies  of other pulsating
  white dwarfs are needed to confirm this value of the axion mass.}
  
\notoc  
  
    
\begin{document}  
  
\maketitle    
  
  
\section{Introduction}  
\label{introduction}  
  
Axions are weakly  interacting hypothetical particles. Their existence  
was proposed  about 35  years ago to  solve the so-called  ``strong CP  
problem''  in  quantum  chromodynamics,   that  is  the  violation  of  
charge-parity  symmetry in strong  interactions \cite{PQ77,We78,Wi78}.  
Comprehensive  accounts  of  the  properties  of axions,  and  of  the  
astrophysical  and  cosmological  searches  aimed at  detecting  these  
elusive particles,  or at least constraining their  properties, can be  
found in  Refs.~\cite{R96,R07,Kim-Carosi}.  Besides its  relevance for  
the standard model,  axions are also natural candidates  to be part of  
the non-baryonic  dark matter of the  Universe. Notwithstanding, their  
contribution to  the non-baryonic dark matter content  of the Universe  
depends on their mass, which  determines the intensity of the coupling  
with matter.  However, the theory that predicts the existence of these  
particles does  not provide any  indication of their mass.   Thus, the  
mass  of  the  axion  has  to  be  determined  either  directly  using  
sophisticated  experimental facilities,  or  indirectly employing  the  
observed  properties  of  astronomical objects  \cite{Vys78,R96}.   In  
particular,  white dwarf  stars ---  see Refs.~\cite{WK08,FB08,review}  
for  recent and  comprehensive  reviews of  their  properties ---  are  
excellent  candidates  to test  the  existence  of weakly  interacting  
particles like axions \cite{primero,R96}.  The reason for this is that  
their evolution is  a simple and well studied  gravothermal process of  
cooling that can be  accurately measured using either their luminosity  
function,  or  by studying  the  secular  variations of  the  periods  of  
variable white dwarfs.  
  
Of the  two types of  axion models proposed  so far ---  the so-called
KVSZ model  \cite{K79,Sea80}, and the DFSZ  model \cite{Dea81,Z80} ---
in  this paper  we will  focus  on the  DFSZ axions,  which couple  to
electrons in addition to photons and hadrons. The coupling strength of
DFSZ axions to electrons is defined through the dimensionless coupling
constant  $g_{\rm ae}$, which  is related  to the  mass of  the axion,
$m_{\rm a}$, through the relation:
  
\begin{equation}  
g_{\rm ae}= 2.8 \times 10^{-14}\  \frac{m_{\rm a} \cos^2 \beta}{1\ {\rm meV}},  
\label{eq1}  
\end{equation}  
  
\noindent where  $\cos^2 \beta$  is a free,  model-dependent parameter
that is  usually set equal  to unity. Thus,  the value of  $m_{\rm a}$
determines how strongly DFSZ axions couple to electrons, and hence how
large  the  axion emissivity  is.   At  the  typical temperatures  and
densities found  in the  cores of white  dwarfs, the emission  of DFSZ
axions takes  place from  the deepest regions  of these  stars, mostly
through bremsstrahlung emission \cite{R86}, the production rate being
\cite{Nea87, Nea88}:
  
\begin{equation}  
\epsilon_{\rm a}= 1.08 \times 10^{23} \frac{g^2_{\rm ae}}{4\pi}   
\frac{Z^2}{A} T_7^4 F(T, \rho)\ \ [{\rm erg/g/s}].   
\label{eq2}  
\end{equation}  
  
Since axions  can (mostly)  freely escape from  the interior  of white  
dwarfs,  their  existence  would  accelerate the  cooling,  with  more  
massive axions  producing a  larger increase in  the rate  of cooling.  
Fortunately,  some   white  dwarfs  pulsate,   and  their  pulsational  
properties depend  sensitively on  the rate of  cooling. Specifically,  
there are  two competing  internal evolutionary processes  that govern  
the rate of change of the pulsation period with time ($\dot\Pi$) for a  
single mode in  pulsating white dwarfs. Cooling of  the star increases  
the periods  as a  result of the  increasing radius of  the degeneracy  
boundary,  whereas residual  gravitational  contraction decreases  the  
periods \citep{Wingetet83}.  Contraction  is still significant for the  
hotter classes of pulsating white  dwarfs, such as the so-called DOVs,  
but it  has been  concluded \citep{Kepleret00} that  for the  class of  
$g$(gravity)-mode pulsating white dwarfs with hydrogen atmospheres ---  
known  as DAVs  or ZZ~Ceti  stars  --- the  evolutionary $\dot\Pi$  is  
dictated solely  by the rate  of cooling.  Therefore, a  comparison of  
the theoretical models of pulsating white dwarfs with the observations  
allows us to  derive useful constraints on  the value of the  mass of the  
axion.  Moreover,   ZZ~Ceti  stars,   with  over  148   members  known  
\cite{Cea10}, provide  us with  a relatively large  observational data  
base to perform such studies.  
  
The star G117$-$B15A is the benchmark member of the ZZ~Ceti class, its  
pulsation  periods  ($\Pi$)  being  215.20~s,  270.46~s  and  304.05~s  
\cite{Kea82}.  The most recent determination  of the rate of change of  
the period at 215~s of this star is $\dot{\Pi}= (4.19 \pm 0.73) \times  
10^{-15}$~s/s \cite{Kea12}.  The possibility of employing the measured  
period drift of the largest  amplitude mode of G117$-$B15A to derive a  
limit on  the mass  of axions  was brought up  almost two  decades ago  
\cite{primero}.   The evolution  of  DA white  dwarf  models with  and  
without axion  emission was considered, and the  theoretical values of  
$\dot{\Pi}$ for increasing masses of  the axion were compared with the  
observed rate of change of period with time of G117$-$B15A.  Employing  
a  semi-analytical treatment,  $m_{\rm a}  \cos^2 \beta=  8.7$~meV    
was obtained.  Subsequently, a detailed  
asteroseismological  model for G117$-$B15A  revealed $m_{\rm  a} \cos^2  
\beta \leq 4.4$~meV \cite{NewA}.  A few years later, an upper limit of  
$13.5 -  26.5$~meV was  derived for the  axion mass using  an improved  
asteroseismological model  for G117$-$B15A, and a  better treatment of  
the  uncertainties  involved  \cite{BKea08}.   Armed with  the  latest  
asteroseismological model \cite{Rea12}  and the most recently measured  
value of $\dot{\Pi}$ \cite{Kea12}, the problem of determining the mass  
of  the  axion  using   white  dwarf  asteroseismology  was  revisited  
\cite{mnras}.   The 215~s  period in  white dwarf  models of  the star  
G117$-$B15A  was  associated with  a  pulsation  mode  trapped in  the  
hydrogen envelope.  The models  strongly indicated the existence of an  
additional  mechanism of energy  loss in  this pulsating  white dwarf,  
consistent with axions of mass $m_{\rm a} \cos^2 \beta \sim 17.4$ meV.  
  
The  star  R548  ---  that  is,  ZZ~Ceti  itself,  with  an  effective  
temperature  $T_{\rm eff} =  11\,990 \pm  200$~K, and  surface gravity  
$\log g= 7.97 \pm 0.05$ \cite{Beea04} --- is another DAV star that has  
been  studied for  the last  few  decades. Both  G117$-$B15A and  R548  
display similar pulsational properties, share periods near $213-215$~s  
and $272-274$~s,  have similar effective temperatures  and masses, and  
are expected to  be similar in structure. Since  the discovery of R548  
\cite{LH71}, there  have been multiple  attempts to measure  the drift  
rate    of     its    pulsation    period     at    $\sim    213.13$~s  
\cite{Sea1980,tomaney,Mea03}. The  rate of change of  this period with  
time has recently been measured  for the very first time \cite{Mea12},  
and forms  our underlying motivation to  determine another independent  
constraint  on the axion  mass by  using the  best asteroseismological  
model for R548 \cite{Rea12}.  
  
The paper is organized as follows.  In Sect.~2 we give a brief account  
of the measurements of the rate  of change of period with time for the  
213.13\,s  period  in  R548,  while  in  Sect.~\ref{asteroseismic}  we  
succinctly  present   the  asteroseismological  model   of  R548.   In  
Sect.~\ref{axion_mass} we  derive a new constraint on  the axion mass.  
Finally,  in  Sect.~\ref{conclusions} we  summarize  our findings  and  
present our concluding remarks.  
  
\section{R548: Measuring the evolutionary cooling rate $\dot{\Pi}$}  
\label{observations}  
  
R548  was the third  DAV star  to be  discovered \cite{LH71}  after HL  
Tau76 \cite{L68} and G44$-$32 \cite{LH69}, but it was chosen to be the  
class prototype as the first  DAV to be deciphered. This star exhibits  
five   independent   pulsation    eigenmodes,   two   of   which   are  
triplets. Their periods and corresponding (amplitudes), measured using  
the  highest  signal-to-noise data  set  from  2007,  are as  follows:  
186.86~s (0.5  mma), 212.77~s (4.3 mma), 212.95~s  (1.2 mma), 213.13~s  
(6.6 mma), 274.25~s (4.1 mma), 274.52~s (1.2 mma), 274.78~s (3.1 mma),  
318.08~s  (0.8 mma),  and 333.64~s  (0.8 mma).   The first  attempt to  
measure the  rate of change of period  with time of R548  was based on  
high-speed     photometry    acquired     from     1970    to     1978  
\cite{Sea1980}. However, only an upper  limit to the drift rate of the  
213.13~s   period  of   $2  \times   10^{-13}$~s/s  could   be  found.  
Nevertheless,  at  that  time  this  measurement  was  a  considerable  
breakthrough, as  R548 was considered  to have the most  stable period  
ever measured in a variable star at visual wavelengths.  
  
The quest to  measure the rate of change of period  with time for R548  
continued for several decades.  Analyzing observations that spanned 31  
years (from 1970 to 2001) an upper limit constraining the evolutionary  
drift   rate   of  the   213~s   period   was   established  in   2003  
\cite{Mea03}. This upper  limit turned out to be  $\dot{\Pi} \leq (5.5  
\pm  1.9)  \times  10^{-15}$~s/s  after including  the  proper  motion  
correction.   A few  years later,  a value  of $(2.1  \pm  1.2) \times  
10^{-15}$~s/s  was  established  \cite{Mea09}  based on  37  years  of  
observations (from 1970 to 2007).  With additional data, the values of  
the drift rate have fluctuated up  and down but the uncertainty of the  
determination has always reduced  monotonically with the square of the  
elongating timebase. Very recently, a reliable measurement of the rate  
of change of period with time for the 213.13~s period in R548 has been  
finally obtained, $\dot{\Pi}=(3.3\pm1.1)\times 10^{-15}$~s/s, using 41  
years  of  time-series  photometry  from 1970  to  2011  \cite{Mea12}.  
During the  last ten  years, the $\dot{\Pi}$  values so  obtained have  
been     consistent     with     the    claimed     measurement     of  
$3.3\times10^{-15}$~s/s  within   uncertainties.   This  improves  our  
confidence in utilizing the measured $\dot{\Pi}$ to constrain the mass  
of the axion.   Additionally, it is worth mentioning  that in the most  
recent  study  of  pulsational   properties  of  R548  \cite{Mea12}  a  
measurement  of the cooling  rate for  any other  period of  the 213~s  
triplet, or for any other mode, has not been claimed. Although in this  
work the $\dot{\Pi}$  values for the 212.78~s period  have 
been determined at each juncture,  these   
values  show  fluctuations  that  are  
significantly larger than  the $1\sigma$ uncertainties.  Consequently,  
in  the absence  of a  reliable determination  of $\dot{\Pi}$  for the  
212.78~s  period, in  this paper  we will  only consider  the measured  
$\dot{\Pi}$ for the 213.13~s.  
  
The  $\dot{\Pi}$  value obtained  for  R548  is completely  consistent  
within uncertainties  with the corresponding  drift rate of the  215 s  
period  of  G117$-$B15A $\dot{\Pi}=(4.19\pm0.73)\times  10^{-15}$~s/s.  
It is also compatible with the  expected cooling of a 
$\sim 0.6 M_{\sun}$ white  
dwarf  harboring  a carbon-oxygen  core.   Consequently, the  213.13~s  
period of  R548 can be used to  derive constraints on the  mass of the  
axion --- see Sect.  \ref{axion_mass}.  
  
\section{Asteroseismological model for R548}  
\label{asteroseismic}  
  
\begin{table}  
\centering  
\caption{Characteristics of R548 derived from a spectroscopic analysis  
  and  results  of  the  best asteroseismological  model.  The  quoted  
  uncertainties  in  the asteroseismological  model  are the  internal  
  errors of our period-fitting procedure.}  
\begin{tabular}{lcc}  
\hline  
\hline  
\noalign{\smallskip}  
 Quantity                        & Spectroscopy      & Asteroseismology                              \\    
\noalign{\smallskip}  
\hline  
\noalign{\smallskip}  
$T_{\rm eff}$ [K]                & $11\,990 \pm 200$ & $11\,627 \pm 390$                             \\  
$M_*/M_{\sun}$                   & $0.590 \pm 0.026$ & $0.609 \pm 0.012$                             \\  
$\log g$                         & $7.97 \pm 0.05$   & $8.03\pm 0.05$                                \\  
$\log(R_*/R_{\sun})$             &    ---            & $-1.904\pm 0.015$                             \\  
$\log(L_*/L_{\sun})$             &    ---            & $-2.594 \pm 0.025$                            \\  
$M_{\rm He}/M_*$                 &    ---            & $2.45 \times 10^{-2}$                         \\  
$M_{\rm H}/M_*$                  &    ---            & $(1.10\pm 0.38) \times 10^{-6}$               \\  
$X_{\rm C},X_{\rm O}$ (center)   &    ---            & $0.26^{+0.22}_{-0.09} , 0.72^{+0.09}_{-0.22}$ \\  
\noalign{\smallskip}  
\hline  
\hline  
\end{tabular}  
\label{table1}  
\end{table}  
  
We have recently performed  a detailed asteroseismological analysis of  
R548 \cite{Rea12} --- among 43 other bright DAVs including G117$-$B15A  
--- using  a very  large grid  of DA  white dwarf  evolutionary models  
characterized by  consistent chemical profiles  for both the  core and  
the envelope, and covering a wide range of stellar masses, thicknesses  
of  the hydrogen envelope  and effective  temperatures.  What  is more  
important, these models were self-consistently generated with the {\tt  
  LPCODE}  evolutionary  code   \cite{Altea05}.   In  particular,  the  
evolutionary calculations were carried  out from the ZAMS, through the  
thermally-pulsing and mass-loss phases on  the AGB, and finally to the  
domain  of   planetary  nebulae  and  white   dwarfs.   The  effective  
temperature, the stellar mass and the mass of the H envelope of our DA  
white dwarf models vary within the ranges $14\,000 \ga T_{\rm eff} \ga  
9\,000$ K, $0.525  \la M_* \la 0.877 M_{\sun}$,  $-9.4 \la \log(M_{\rm  
H}/M_*) \la -3.6$,  where the value of the upper  limit of $M_{\rm H}$  
depends on $M_*$ and is fixed by the prior evolution.  For the sake of  
simplicity, and also  for coherence, the mass of  the helium layer was  
kept fixed at the value predicted by the evolutionary calculations for  
each sequence \cite{Rea12}.  
  
The theoretical  periods were assessed by means  of a state-of-the-art  
pulsation code \cite{CA06}.  To  find an asteroseismological model for  
R548, we searched for the model that minimizes a quality function that  
measures the distance between  the theoretical ($\Pi^{\rm t}$) and the  
observed  ($\Pi^{\rm  o}$) periods  \cite{Rea12}.   A single  best-fit  
model with the characteristics  shown in Table~\ref{table1} was found.  
The second column of Table~\ref{table1} contains the spectroscopically  
determined values of $T_{\rm eff}$ and $\log g$ of R548 \cite{Beea04},  
and the stellar mass  \cite{Rea12}.  The parameters characterizing the  
asteroseismological    model   are    shown   in    column    3.    In  
Table~\ref{table2} we compare the observed and the theoretical periods  
--- along  with  the corresponding  mode  identification  --- and  the  
expected  rates of  change  of the  periods  with time.   It is  worth  
mentioning  that the  model  nearly reproduces  the observed  periods,  
although  the  modes with  periods  at  318.08~s  and 333.64~s  remain  
relatively   poorly   matched.     The   most   relevant   result   of  
Table~\ref{table2} is  that the observed  rate of change of  the 213~s  
period ($3.3 \times  10^{-15}$~s/s) is $\sim 3$ times  larger than the  
theoretically expected  value ($1.08 \times  10^{-15}$~s/s). Since the  
rate  of  change  of  period  with  time of  this  mode  reflects  the  
evolutionary timescale of the  star, then the disagreement between the  
observed and  theoretical values of $\dot\Pi$ is  suggestive that R548  
could  be  cooling faster  than  the  cooling  rate predicted  by  the  
standard theory of white dwarf evolution.  
  
\begin{table}  
\centering  
\caption{The observed and theoretical  periods of R548, along with the  
  corresponding  mode  identification,  and  the subsequent  rates  of  
  change of  period with  time (computed without including axion cooling). 
  For the  213~s and 274~s  triplets, we  
  only list the periods corresponding to the central component.}  
\begin{tabular}{cccccc}  
\hline  
\hline  
\noalign{\smallskip}  
$\Pi^{\rm o}$        & $\Pi^{\rm t}$       & $\ell$ & $k$ & $\dot{\Pi}^{\rm o}$  & $\dot{\Pi}^{\rm t}$\\  
\noalign{\smallskip}  
$[$s$]$              &  $[$s$]$            &        &    & $[10^{-15} $s/s$]$   & $[10^{-15}  $s/s$]$\\    
\noalign{\smallskip}  
\hline  
\noalign{\smallskip}  
186.86               &   187.59            & 1      &  1  & ---           & 2.51 \\  
212.95               &   213.40            & 1      &  2  & $3.3 \pm 1.1$ & 1.08 \\  
274.52               &   272.26            & 1      &  3  & ---           & 3.76 \\  
318.08               &   311.36            & 2      &  8  & ---           & 6.32 \\  
333.64               &   336.50            & 2      &  9  & ---           & 8.80 \\  
\noalign{\smallskip}  
\hline  
\hline  
\end{tabular}  
\label{table2}  
\end{table}  
  
We note that in a previous work \cite{BKea08b} it was found an average  
rate  of change  of the  period with  time of  $(2.91\pm  0.29) \times  
10^{-15}$~s/s   for  the   213~s  period   of  R548.    However,  this  
determination was based on identifying  the 213~s period with a radial  
order   $k=1$  mode,  instead   of  $k=2$,   which  is   our  best-fit  
solution. Similar  to our  findings for the  case of  G117$-$B15A, the  
rate of change of period with time for the $\ell=1$, $k=2$ mode in our  
asteroseismological model is substantially smaller than for the dipole  
modes  with $k=1$  and 3,  and much  smaller than  for  the quadrupole  
($\ell=2$) modes with $k=8$ and 9.   We expect that the $k=2$ mode has  
the  smallest $\dot\Pi$  because it  is a  mode trapped  in  the outer  
hydrogen envelope in  our model white dwarf. This  also applies to the  
215~s  mode of  G117$-$B15A \cite{mnras}.   Mode trapping  reduces the  
rate of change  of period with time by  up to a factor of  $\sim 3$ if  
the  mode   is  trapped  in  the  outer   hydrogen  envelope,  because  
gravitational  contraction  --- that  is  still  appreciable in  these  
regions  ---   reduces  the  net  increase  in   period  from  cooling  
\cite{B96}.    Since  the   $k=2$   mode  is   somewhat  affected   by  
gravitational contraction,  it is  less sensitive to  the evolutionary  
cooling.   However, the  change of  the period  due to  the increasing  
degeneracy resulting from cooling is  still larger than the change due  
to residual  contraction, and so, $\dot{\Pi}  > 0$.  As  a result, the  
period of the $k=2$ mode is  still sensitive to cooling and will allow  
us to constrain the mass of the axion.  
  
An important  point in our  analysis is to estimate  the uncertainties  
affecting  the  value   of  $\dot\Pi$  for  the  $k=2$   mode  in  our  
asteroseismological  model,  because   they  directly  translate  into  
uncertainties in the derived axion  mass. Here, we will adopt the same  
approach we used previously  for G117$-$B15A \cite{mnras}. We estimate  
an uncertainty $\sim 0.03 \times 10^{-15}$~s/s for $\dot{\Pi}^{\rm t}$  
of   the   213~s   period    due   to   the   uncertainties   in   the  
$^{12}$C$(\alpha,\gamma)^{16}$O reaction rate. Since the $k=2$ mode is  
trapped in the hydrogen envelope of our model white dwarf, the precise  
abundances  of carbon  and oxygen  in  the core  do not  significantly  
affect the calculated  period and the corresponding rate  of change of  
period with time.  We also  estimate an uncertainty in the theoretical  
$\dot\Pi$  of $\sim  0.06  \times 10^{-15}$~s/s  due  to the  internal  
errors in  fitting the  period.  Fortunately, these  uncertainties are  
small and do not contribute  significantly to the uncertainties in the  
derived axion mass.  As shown below, the uncertainties in the inferred  
value  of  $m_{\rm a}$  are  dominated  by  the uncertainties  in  the  
observed rate of change of period with time of the 213~s period.  
  
\section{Axion emission and inference of the axion mass}  
\label{axion_mass}  
  
\begin{figure}  
\centering  
\includegraphics[clip, width=0.8\textwidth, angle=0]{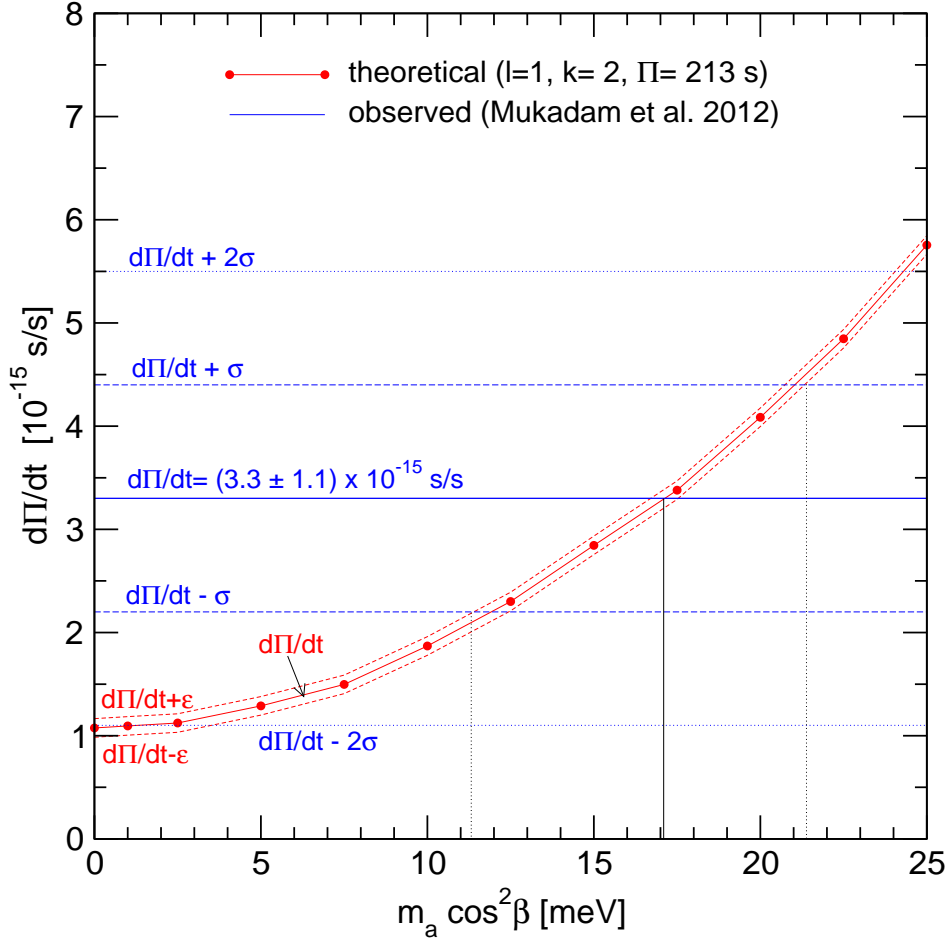}  
\caption{The rate of period change for  the mode with $\ell = 1$, $k =  
  2$ and period $\sim 213$~s of our asteroseismological model for R548  
  ($T_{\rm eff}= 11\,627  \pm 390$ K) as a function  of the axion mass  
  (solid red curve  with dots). Dashed curves represent  the errors in  
  $\dot{\Pi}$ due to internal uncertainties in the modeling and in the  
  asteroseismological  procedure. The  horizontal  lines indicate  the  
  observed value  with its corresponding uncertainties in  the case of  
  the solution  with a non-linear least squares  fit, $\dot{\Pi}= (3.3  
  \pm 1.1) \times 10^{-15}$~s/s.}  
\label{figure1}  
\end{figure}  
  
We have found that the theoretically expected rate of change of period  
with time  for the $k=  2$ mode is  distinctly smaller than  the value  
measured for R548, suggesting the existence of some additional cooling  
mechanism in this star.  Here,  we assume that this additional cooling  
can be entirely  attributed to the emission of  axions.  Similar to it  
was done in our previous  papers \cite{NewA,mnras}, we have computed a  
set of DA white dwarf  cooling sequences incorporating the emission of  
axions. This has been done  considering different axion masses and the  
same   structural    parameters   ($M_*$,   $M_{\rm    H}$)   as   the  
asteroseismological  model in  Table~1.  We  have adopted  a  range of  
values for the  mass of the axion $0 \leq m_{\rm  a} \cos^2 \beta \leq  
30$~meV, and  also employed the  most up-to-date axion  emission rates  
\cite{Nea87,Nea88}.   The   evolutionary  calculations  including  the  
emission of axions were started at evolutionary stages long before the  
ZZ~Ceti phase to  ensure that the cumulative effect  of axion emission  
has reached an equilibrium value before the models reach the effective  
temperature of R548.  
  
We found that the pulsation periods  for the modes with $\ell= 1,\ k =  
1, 2,  3$, and $\ell=  2,\ k= 8,  9$ of the  asteroseismological model  
experience negligible variations with increasing values of the mass of  
the axion,  $m_{\rm a}$.  This  is because the periods  themselves are  
not appreciably affected by the  small changes in the structure of the  
white  dwarf produced  by the  emission of  axions.  In  contrast, the  
rates of the change of period with time are strongly affected by axion  
emission,  substantially increasing for  increasing values  of $m_{\rm  
  a}$.  In particular, for the  mode of interest ($\ell=1,\ k= 2$) the  
rate of  change of period  increases by a  factor of order of  10 when  
$m_a$  goes from 0  to $30$  meV. This  convincingly shows  that, even  
though this mode is less  sensitive to the evolutionary cooling of the  
star compared to the modes with  $k=1$ and $k=3$, it is still a useful  
tool to constrain the mass of the axion.  
  
In Fig.~\ref{figure1} we display  the theoretical value of $\dot{\Pi}$
corresponding to  the period $\Pi=213$~s for increasing  values of the
axion mass (red  solid curve).  The dashed curves  embracing the solid
curve  represent   the  uncertainty   in  the  theoretical   value  of
$\dot{\Pi}$, $\varepsilon_{\dot{\Pi}}= 0.09 \times 10^{-15}$~s/s. This
value has been obtained  considering the uncertainty introduced by our
lack  of  precise  knowledge  of  the  $^{12}$C$(\alpha,\gamma)^{16}$O
reaction  rate  and  by  the  internal uncertainties  in  fitting  the
asteroseismological model.  Additionally,  we assume that both sources
of  uncertainties  do  not depend  on  the  mass  of the  axion.  The
horizontal  (blue)   solid  line  indicates  the   observed  value  of
$3.3\times  10^{-15}$~s/s,  whilst  its  corresponding  $1\sigma$  and
$2\sigma$ uncertainties \cite{Mea12} are shown using dashed and dotted
lines, respectively.  If one standard deviation from the observational
value is considered, we conclude that the mass of the axion is $m_{\rm
a} \cos^2 \beta= \left(17.1^{+4.3}_{-5.8} \right)$~meV.  This value is
in complete agreement with  the axion mass inferred using G117$-$B15A,
$m_{\rm   a}  \cos^2   \beta=   \left(17.4^{+2.3}_{-2.7}  \right)$~meV
\cite{mnras}. Note,  however, that within $2\sigma$,  our results are
compatible with $m_{\rm a}=0$.

\section{Discussion and conclusions}  
\label{conclusions}  
  
In this paper we have derived an improved value of the mass of axions,
assuming  that the  enhanced rate  of cooling  of the  pulsating white
dwarf R548 is entirely due to  the emission of axions.  In doing so we
have employed  a detailed asteroseismological  model for R548  --- the
prototype  of  the DAV  stars  ---  obtained  using full  evolutionary
calculations of DA white  dwarf models \cite{Rea12}.  Our calculations
used the  recent determination  of the rate  of change of  period with
time for  the largest  amplitude mode of  this star  \cite{Mea12}.  We
found  that if  the 213~s  period in  R548 is  associated with  a mode
trapped  in  the  H  envelope,  our theoretical  models  indicate  the
existence of  an additional cooling mechanism in  this pulsating white
dwarf,  consistent  with  axions  of  mass $m_{\rm  a}  \cos^2  \beta=
\left(17.1^{+4.3}_{-5.8}  \right)$~meV  at  the  $1\sigma$  level,  or
$m_{\rm a} \cos^2  \beta=\left(17.1^{+7.2}_{-17.1} \right)$~meV at the
$2\sigma$  level.  Equivalently,  in terms  of the  constant coupling,
this mass  of the axion  can be expressed  as $g_{\rm ae}=  4.8 \times
10^{-13}$.  Our  value for  the axion mass  is in  excellent agreement
with  that inferred from  the pulsating  white dwarf  G117$-$B15A, but
considerably  larger than  that obtained  from the  hot branch  of the
white    dwarf   luminosity    function,    $m_{\rm   a}\sim    5$~meV
\cite{wdlf,IEA09}.  Indeed,  a detailed analysis of the  hot branch of
the white  dwarf luminosity function suggests that  values larger than
$10$~meV can be safely excluded.   However, at the $2\sigma$ level our
determination is  consistent with that obtained using  the white dwarf
luminosity function.  Moreover, both techniques agree  in finding that
an anomalous rate of cooling  of white dwarfs in this luminosity range
exists. If this  anomalous rate of cooling can  be entirely attributed
to  the  emission  of   axions  deserves  further  scrutiny.   It  is,
nevertheless, worth stressing that  both methods are complementary and
equally sensitive to the emission  of axions in white dwarfs, and that
both suggest that axions  do exist, with a mass on the  order of a few
meV.
  
Another  bound  on  the  axion-electron  coupling  comes  from  helium
ignition in low-mass stars on the red giant branch.  Using this method
an upper  limit $m_{\rm a}  \cos^2 \beta \lesssim 9$~meV  was obtained
\cite{RW1995}.  The axion mass  derived from pulsating white dwarfs is
substantially larger than  this upper limit, and axions  with the mass
derived in  this paper would modify  the tip of the  red giant branch,
although the uncertainties involved in the method of \cite{RW1995} are
comparable  to those  affecting our  procedure. Finally,  it  is worth
mentioning that axions with masses in the range of values derived here
would  provide  a  strong   energy  loss  channel  for  core  collapse
supernovae  and neutron  stars  \cite{2011PhRvD, 2012arXiv1205.6940K}.
However, we emphasize that again our results are compatible with these
upper bounds at the $2\sigma$ level.
  
With the inference of the axion mass from R548 reported in this paper,  
we now  have two independent determinations  of the mass  of the axion  
from  two different  pulsating  white dwarfs,  which  agree with  each  
other.  These determinations imply  that axions couple with electrons,  
and therefore  must be of the  DFSZ type.  Although  these results are  
encouraging, it  is worth considering that both  stars have comparable  
observational properties  --- with similar  effective temperatures and  
gravities --- and also share very similar pulsational characteristics,  
i.e., the periods  of their dominant modes. In  addition, the rates of  
change of  period with time used to  derive the mass of  the axion are  
associated with the same $\ell=1,\ k= 2$ pulsation mode in both stars.  
Therefore, we can expect to yield the same value  of the mass of  
the axion based on the $\dot\Pi$  of the same eigenmode in two similar  
stars.  Hence, it would be desirable to have a measurement of the rate  
of change  of period with time  for another DAV  star with pulsational  
and  spectroscopic  characteristics   very  different  from  those  of  
G117$-$B15A  and  R548.  Examples  of  such  white  dwarfs  are  L19-2  
\cite{OW87} and G226$-$29.  Also, a  measurement of the rate of change  
of period with time for white dwarfs with hydrogen deficient envelopes  
--- like EC20058$-$5234 \cite{Dea10}  and KIC~8626021 \cite{Oea11} ---  
will allow us to obtain another independent bound on the axion mass.  
  
\acknowledgments Part  of this work  was supported by  AGENCIA through  
the Programa  de Modernizaci\'on Tecnol\'ogica BID  1728/OC-AR, by the  
PIP   112-200801-00940   grant    from   CONICET,   by   MCINN   grant  
AYA2011--23102, by the ESF EUROCORES Program EuroGENESIS (MICINN grant  
EUI2009-04170),  by  the  European  Union  FEDER  funds,  and  by  the  
AGAUR.  A.S.M.  acknowledges  NSF  for the  grant  AST-1008734.   This  
research has made use of NASA's Astrophysics Data System.  
  
\bibliographystyle{JHEP}  
\bibliography{R548-r1}  
  
\end{document}